\documentclass[prl,10pt,showkeys,showpacs,twocolumn]{revtex4}

\usepackage[dvips]{graphics}

\begin{document}

\title{Reply to ``Comment on `Entropy of classical systems with long-range interactions' ''}

\author{T.\ M.\ Rocha Filho, A.\ Figueiredo \& M.\ A.\ Amato}
\affiliation{Instituto de F\'\i{}sica, Universidade de
Bras\'\i{}lia\\ CP 04455, 70919-970 - Bras\'\i{}lia, Brazil}



\maketitle

In the comment on our recent letter~\cite{r1} Rapisarda and coworkers probably missed the relevant
points.
Let us start by those summarized at the end of their comment as being erroneous statements in our letter:
(i) {\it In the limit $N\rightarrow\infty$ the interparticle correlations are
negligible is untrue for long range correlations}.
This statement is simply not correct. It is well known in the literature
that particles are uncorrelated in this limit and a mean-field description
by the Vlasov Equation (VE) is exacti~\cite{r2,r3}. In particular see eq.~(18) of Ref.~\cite{r3}.
(ii) {\it The supposition that all microstates compatible with the given constraints are equally probable
is invalid, even for the microcanonical ensemble... strong indications of nonergodicity are available
in the literature}. Our letter shows undoubtly that this supposition
leads to the correct results obtained from other procedures.
Our approach leads to non-equiprobable states arising,
despite the use of the Boltzmann-Gibbs (BG) entropy, from the non-linear Casimir constrains of
the VE. The BG entropy implies ergodicity in the constant energy hypersurface in phase space only
if there is no constraints other than energy and norm.
Unfortunately Rapisard et al.\ are quite vague on this point. It is properly addresses and thoroughly
discussed in Ref.~\cite{r7}.
(iii) {\it The BG entropy is then the correct form to be used
is trivially correct for $N\rightarrow\infty$ after $t\rightarrow\infty$, and clearly wrong the other way around,
since the distribution of velocities is not Maxwellian in the quasi-stationary state}. The logic argumentation
is incomplete as can be perceived by reading our letter. The maximization of the BG entropy
with the Casimir constraints leads to a non-Maxwellian distribution function, as we explicitly stated.

Our reasoning is not restricted to homogeneous systems.
We have shown that stationary states of the VE,
which correspond to statistical stationary states of {\it long-range classical hamiltonian systems}, are {\it always}
the same as those obtained from the maximization of the BG entropy with the VE constraints.
It is also important to note that there is no $H$-Theorem for the VE, since it is essentially
reversible. Therefore there is no guarantee that a system with long-range interaction settles down in
some stationary state, which explains results from Reference~\cite{nv1} (reference [8] of the comment).
This is in fact expected from the mean field attribute of the model, described
by the VE, and cannot be explained by Tsallis approach.
Also the results in Ref.\ [4] of the
comment on the generalized central limit theorem are not applicable to the present case. For a
less restrictive and algebraic method for the central limit theorem see Ref.~\cite{r6}.
If the system has a finite number of particles, the right hand side of the VE must me modified,
leading to a collisional term responsible for the finite life-time of the quasi-stationary states. We don't
know if they meant that TE is not valid in the $N\rightarrow\infty$ limit (after $t\rightarrow\infty$
of course), but {\it finite-size effects}
are small unless $N$ itself is really small. Therefore we cannot understand how these very small corrections are
capable of transforming an entropy form depending on an infinite number of parameters into the TE,
which depends only on two parameters. Besides it was shown that finite systems have stationary states not
compatible with the Tsallis statistics.

Non-extensivity is a direct consequence of correlations in the system, without resorting
to any different entropy definition or any new kinetic theory~\cite{r8}. If two systems
are not statistically correlated then the total entropy must be the sum of the entropy of
each system. This is required if one wishes to have a consistent termodynamical interpretation
of entropy, and to avoid the type of condradictions raised by Nauenberg~\cite{r9} (Tsallis reply
to Nauenberg~\cite{r10} is far from satisfactory).
Tsallis informational entropy has no physical meaning in the frame of statistical mechanics.

\end{document}